\documentclass[final,3p,times,twocolumn]{elsarticle}
\usepackage{graphicx}
\usepackage{dcolumn}
\biboptions{square,comma,sort&compress}

\journal{Physics Letters B}

\begin{document}

\begin{frontmatter}

\title{Quadrupole Moments of Collective Structures up to Spin $\sim$ $65\hbar$ in $^{157}$Er and $^{158}$Er: A Challenge for Understanding Triaxiality in Nuclei}

\author[FSU]{X. Wang\fnref{fn1}}
\author[FSU]{M. A. Riley}
\author[DSL]{J. Simpson}
\author[LPU]{E. S. Paul}
\author[DSL]{J. Ollier}
\author[ANLP]{R. V. F. Janssens}
\author[UND]{A. D. Ayangeakaa}
\author[LPU]{H. C. Boston}
\author[ANLP]{M. P. Carpenter}
\author[ANLP,UMD]{C. J. Chiara}
\author[UND]{U. Garg}
\author[UNA]{D. J. Hartley}
\author[LPU]{D. S. Judson}
\author[ANLN]{F. G. Kondev}
\author[ANLP]{T. Lauritsen}
\author[MNU]{N. M. Lumley}
\author[UND]{J. Matta}
\author[LPU]{P. J. Nolan}
\author[LBL]{M. Petri}
\author[LPU]{J. P. Revill}
\author[UTK]{L. L. Riedinger}
\author[LPU]{S. V. Rigby}
\author[LPU]{C. Unsworth}
\author[ANLP]{S. Zhu}
\author[LIT]{I. Ragnarsson}

\fntext[fn1]{Corresponding author; e-mail: xwang3@nucmar.physics.fsu.edu}

\address[FSU]{Department of Physics, Florida State University, Tallahassee, FL 32306, USA}
\address[DSL]{STFC Daresbury Laboratory, Daresbury, Warrington, WA4 4AD, United Kingdom}
\address[LPU]{Department of Physics, University of Liverpool, Liverpool, L69 7ZE, United Kingdom}
\address[ANLP]{Physics Division, Argonne National Laboratory, Argonne, IL 60439, USA}
\address[UND]{Physics Department, University of Notre Dame, Notre Dame, IN 46556, USA}
\address[UMD]{Department of Chemistry and Biochemistry, University of Maryland, College Park, MD 20742, USA}
\address[UNA]{Department of Physics, United States Naval Academy, Annapolis, MD 21402, USA}
\address[ANLN]{Nuclear Engineering Division, Argonne National Laboratory, Argonne, IL 60439, USA}
\address[MNU]{Schuster Laboratory, University of Manchester, Manchester, M13 9PL, United Kingdom}
\address[LBL]{Nuclear Science Division, Lawrence Berkeley National Laboratory, Berkeley, CA 94720, USA}
\address[UTK]{Department of Physics and Astronomy, University of Tennessee, Knoxville, TN 37996, USA}
\address[LIT]{Division of Mathematical Physics, LTH, Lund University, P. O. Box 118, SE-221 00 Lund, Sweden}

\begin{abstract}
The transition quadrupole moments, $Q_{\rm t}$, of four weakly populated collective bands 
up to spin $\sim$ $65\hbar$ in $^{157,158}$Er have been measured to be ${\sim}11~{\rm eb}$ 
demonstrating that these sequences are associated with large deformations. 
However, the data are inconsistent with calculated values from cranked Nilsson-Strutinsky 
calculations that predict the lowest energy triaxial shape to be associated with rotation 
about the short principal axis. The data appear to favor either a stable triaxial shape 
rotating about the intermediate axis or, alternatively, a triaxial shape with larger deformation 
rotating about the short axis. These new results challenge the present understanding of triaxiality in nuclei. 
\end{abstract}


\end{frontmatter}

The question of the occurrence of asymmetric shapes in nature is a general one touching on diverse systems 
spanning many length scales, for example, large galaxies (diameter ${\sim}10^{20}~{\rm m}$), 
planets (${\sim}10^{8}-10^{9}~{\rm m}$), asteroids (${\sim}10^{5}-10^{6}~{\rm m}$), and small 
red blood cells (${\sim}10^{-5}~{\rm m}$) or atomic clusters (${\sim}10^{-9}-10^{-8}~{\rm m}$). 
In nuclear physics, the occurrence of stable asymmetric or triaxial shapes (${\sim}10^{-14}~{\rm m}$) is 
a long-standing prediction of theory~\cite{Bohr-75-book}, and had been sought for experimentally for decades. 
However, it was not until 2001 that compelling evidence for nuclei with a robust 
triaxial shape was revealed in the rare-earth region ($A{\sim}165$) through the observation of rotational 
structures associated with the unique ``wobbling'' excitation mode~\cite{Odegard-PRL-86-5866-01}. 
A triaxial nuclear shape, which has distinct short, intermediate, 
and long principal axes, is commonly described using the parameters ($\varepsilon_2$, $\gamma$) of the 
Lund convention~\cite{Andersson-NPA-268-205-76}, where $\varepsilon_2$ and $\gamma$ represent the eccentricity 
from sphericity and triaxiality, respectively. At high spin, collective rotation about the short axis, 
corresponding to a positive $\gamma$ value ($0^{\circ}<{\gamma}<60^{\circ}$), usually has the lowest excitation 
energy based on moment of inertia considerations~\cite{Szymanski-83-book,Carlsson-PRC-78-034316-08}. 
Thus, this mode is expected to be favored over rotation about the intermediate axis ($-60^{\circ}<{\gamma}<0^{\circ}$). 

Recently, four rotational bands with high moments of inertia were observed at angular momenta up to $65{\hbar}$ in 
$^{157,158}$Er~\cite{Paul-PRL-98-012501-07}. These sequences bypass the energetically favored ``band-terminating'' 
states~\cite{Tjom-PRL-55-2405-85,Simpson-PLB-327-187-94} and extend discrete $\gamma$-ray 
spectroscopy to the so-called ``ultrahigh-spin regime'' ($I=50-70$). 
Cranking calculations~\cite{Bengtsson-PSC-T5-165-83,Dudek-PRC-31-298-85} had predicted collective configurations 
at large deformation to become competitive in energy for spins above $50\hbar$. Configurations with 
$\varepsilon_{2}\sim{0.34}$ and a positive value of $\gamma{=}20^{\circ}-25^{\circ}$ were 
predicted to be low in energy in Ref.~\cite{Dudek-PRC-31-298-85} and re-confirmed in the calculations reported in 
Ref.~\cite{Paul-PRL-98-012501-07}. Thus, this shape minimum had been suggested to correspond to the ultrahigh-spin bands 
in $^{157,158}$Er~\cite{Paul-PRL-98-012501-07}. This suggestion provided an opportunity to test the cranking model, 
a description widely employed in high-spin nuclear physics in the 
limit of extreme angular momentum. Hence, with the purpose of investigating the deformation associated 
with these structures at ultrahigh spin and of elucidating their character, 
a DSAM experiment was performed to extract transition quadrupole moments. 

In this letter, the transition quadrupole moments ($Q_{\rm t}$) of the collective bands with high moments of inertia, 
bands 1 and 2 (adopting the labelling of Ref.~\cite{Paul-PRL-98-012501-07}), in $^{157,158}$Er are reported. 
In comparison with the calculations mentioned above and new calculations using the 
cranked Nilsson-Strutinsky (CNS) model~\cite{Carlsson-PRC-74-R011302-06}, 
these results appear to be inconsistent with the rotation of the theoretically favored triaxial shape. 
The implications of this rather surprising observation are discussed. 

\begin{figure}
\begin{center}
\includegraphics[angle=0,width=0.86\columnwidth]{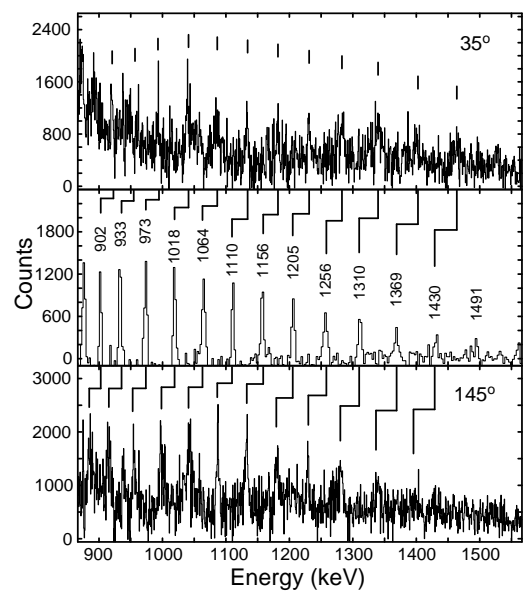}
\caption{Double-gated summed coincidence spectra from the present data for band 1 in $^{158}$Er at two 
typical angles, $35^{\circ}$ (top panel) and $145^{\circ}$ (bottom) with respect to the beam direction. 
The triple-gated summed coincidence spectrum from the thin-target data~\cite{Paul-PRL-98-012501-07} for the same band 
is shown as a reference in the middle panel. The energy values of the inband transitions (in keV) 
are given in the thin-target spectrum, while the peak positions of shifted $\gamma$ rays observed at the 
two angles are marked by lines in the other two spectra.\label{fig:158Er_band1_2gt}}
\end{center}
\end{figure}

The lifetime measurement was carried out using the Doppler Shift Attenuation Method (DSAM)~\cite{Devons-PPSA-68-18-55}. 
A 215 MeV $^{48}$Ca beam was delivered by the ATLAS facility at Argonne National Laboratory and bombarded a 1 mg/cm$^{2}$ $^{114}$Cd 
target backed by a 13 mg/cm$^{2}$ $^{197}$Au layer. A 0.07 mg/cm$^{2}$ $^{27}$Al layer between Cd and Au was used 
to prevent the migration of the target material into the backing. The emitted $\gamma$ rays were detected 
by the Gammasphere spectrometer~\cite{Janssens-NPN-6-9-96}, which consisted of 101 Compton-suppressed HPGe detectors. 
A total of $9.9\times{10}^{9}$ events were accumulated, each one containing at least four coincident $\gamma$ rays. 
Due to the low intensity of the bands of interest, ${\sim}10^{-4}$ of the respective channel intensity, 
several techniques~\cite{Fischer-PRC-54-R2806-96,Wang-PRC-163Tm-07} were utilized in the data analysis. 
An initial analysis using the RadWare software package~\cite{Radford-NIMA-361-306-95} 
proved beneficial before sorting the data into a BLUE database~\cite{Cromaz-NIMA-462-519-01} 
to generate background-subtracted coincidence spectra at different detector angles~\cite{Starosta-NIMA-515-771-03}. 
Examples of the resulting spectra at selected angles can be found in Fig.~\ref{fig:158Er_band1_2gt}. 

\begin{figure}
\begin{center}
\includegraphics[angle=0,width=1.0\columnwidth]{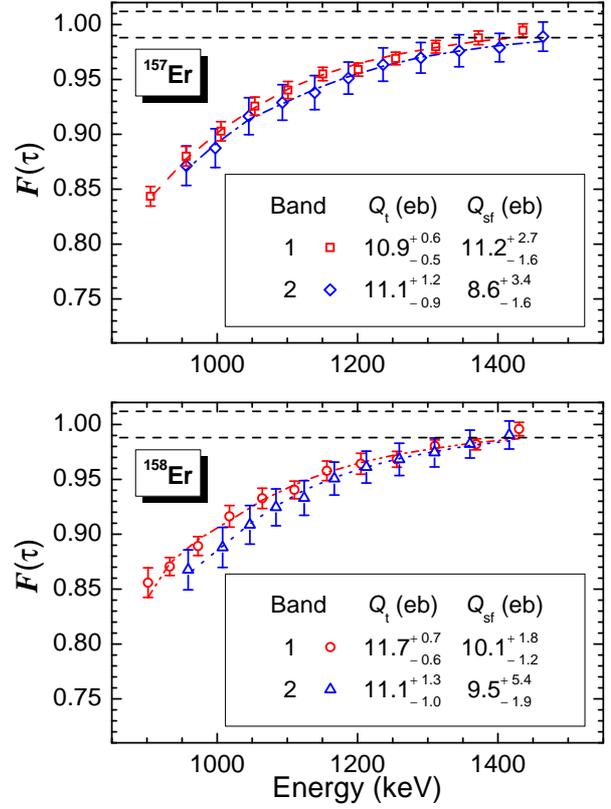}
\caption{(Color online) Measured $F(\tau)$ values as a function of the $\gamma$-ray energy with best-fit 
curves for the four bands in $^{157}$Er and $^{158}$Er. The two horizontal dashed lines in each panel 
show the range of initial recoil velocities of the Er nuclei within the $^{114}$Cd target layer. 
The insets summarize the measured quadrupole moments. The error bars are statistical only, $\it{i.e.}$, they do not include 
the ${\sim}15\%$ error associated with the systematic uncertainty in the stopping powers~\cite{Broude-73-book} 
(see text).\label{fig:157-158Er_TSD_ftau_fit}}
\end{center}
\end{figure}

The fractional Doppler shifts $F(\tau)$ and the associated errors for the inband transitions covering 
an estimated spin range of $30-60{\hbar}$~\cite{Paul-PRL-98-012501-07} were subsequently extracted from 
linear fits of the energy shifts as a function of detector angle $\theta$. 
The resulting $F(\tau)$ values for the four bands are presented in Fig.~\ref{fig:157-158Er_TSD_ftau_fit}. 
The larger $F(\tau)$ uncertainties of the two band 2 sequences relate to their weaker intensities 
as compared with the bands 1. The fit of the $F(\tau)$ values was performed using the computer code 
MLIFETIME~\cite{Moore-unpublsh}, which uses stopping powers provided by the SRIM 2003 code~\cite{Ziegler-85-book}. 
It is worth mentioning two of the assumptions made in the analysis: 
(1) all levels in a band have the same transition quadrupole moment, $Q_{\rm t}$; and 
(2) the sidefeeding into each level in a band is modeled as a single cascade with a common, constant 
quadrupole moment, $Q_{\rm sf}$, and characterized by the same dynamic moment of inertia as the main band 
into which it feeds. 
A ${\chi}^2$ minimization with the parameters $Q_{\rm t}$ and $Q_{\rm sf}$ was then performed to the measured $F(\tau)$ 
values for each of the four bands. This commonly adopted procedure has proved to be reliable 
for SD and TSD bands in this region, see Refs.~\cite{Nisius-PLB-392-18-97,Amro-PLB-506-39-01}, for example. 
Details of the fitting process can be found in Refs.~\cite{Wang-07-thesis,Wang-PRC-163Tm-07}. 
The resulting transition quadrupole moments, $Q_{\rm t}$, and side-feeding quadrupole moments, $Q_{\rm sf}$, 
are summarized in the insets of Fig.~\ref{fig:157-158Er_TSD_ftau_fit}. For comparison, the ground-state band 
of $^{158}$Er has a measured $Q_{\rm t}$ of ${\sim}6~{\rm eb}$ (for states up to spin $20\hbar$)~\cite{Shepherd-PRC-65-034320-02} 
and is associated with a normal prolate shape ($\varepsilon_{2}\sim{0.2}$, $\gamma{=}0^{\circ}$). 
Thus, these four bands with measured $Q_{\rm t}{\sim}11~{\rm eb}$ are associated with a large deformation. 

\begin{figure}
\begin{center}
\includegraphics[angle=0,width=0.8\columnwidth]{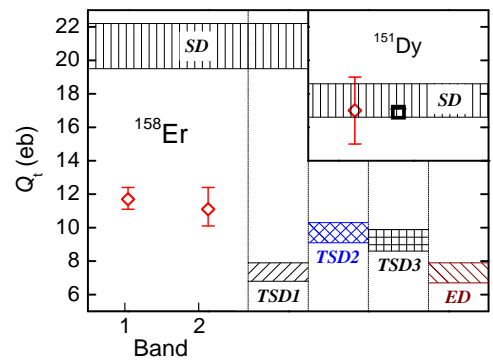}
\caption{(Color online) Measured transition quadrupole moments ($Q_{\rm t}$) of bands 1 and 2 
in $^{158}$Er and of the yrast SD band in $^{151}$Dy (inset), compared with the predicted $Q_{\rm t}$ values 
associated with the minima of interest. The experimental results are plotted with diamonds for the present work 
and a square for Ref.~\cite{Nisius-PLB-392-18-97}. The calculated CNS values in the spin range $30-60{\hbar}$ 
are illustrated as horizontal shaded areas with the heights proportional to their ranges. 
See text for details.\label{fig:Qt_exp_cal_comp}}
\end{center}
\end{figure}

It is well known that an additional systematic error of ${\sim}15\%$ needs to be added for $Q_{\rm t}$ 
measurements with the DSAM technique~\cite{Broude-73-book}. However, another measurement, focussing 
on $^{154}$Er, was performed under very similar experimental conditions~\cite{Revill-to-be-published}, 
where the $Q_{\rm t}$ of the yrast superdeformed band in $^{151}$Dy was extracted. The measured 
value of ${17}{\pm}2~{\rm eb}$ agrees with the reported one of $16.9^{+0.2}_{-0.3}~{\rm eb}$ 
of Ref.~\cite{Nisius-PLB-392-18-97} and is reproduced well by the CNS calculations for a prolate superdeformed shape 
(see Fig.~\ref{fig:Qt_exp_cal_comp}). Hence, this comparison provides a consistency check for the 
experimental approach and the stopping powers used in the present data analysis. This $^{151}$Dy $Q_{\rm t}$ 
value can also be seen in Fig.~\ref{fig:Qt_exp_cal_comp}, where the data for the $^{158}$Er bands are displayed 
and a comparison with CNS results is provided. 

\begin{figure}
\begin{center}
\includegraphics[angle=0,width=0.86\columnwidth]{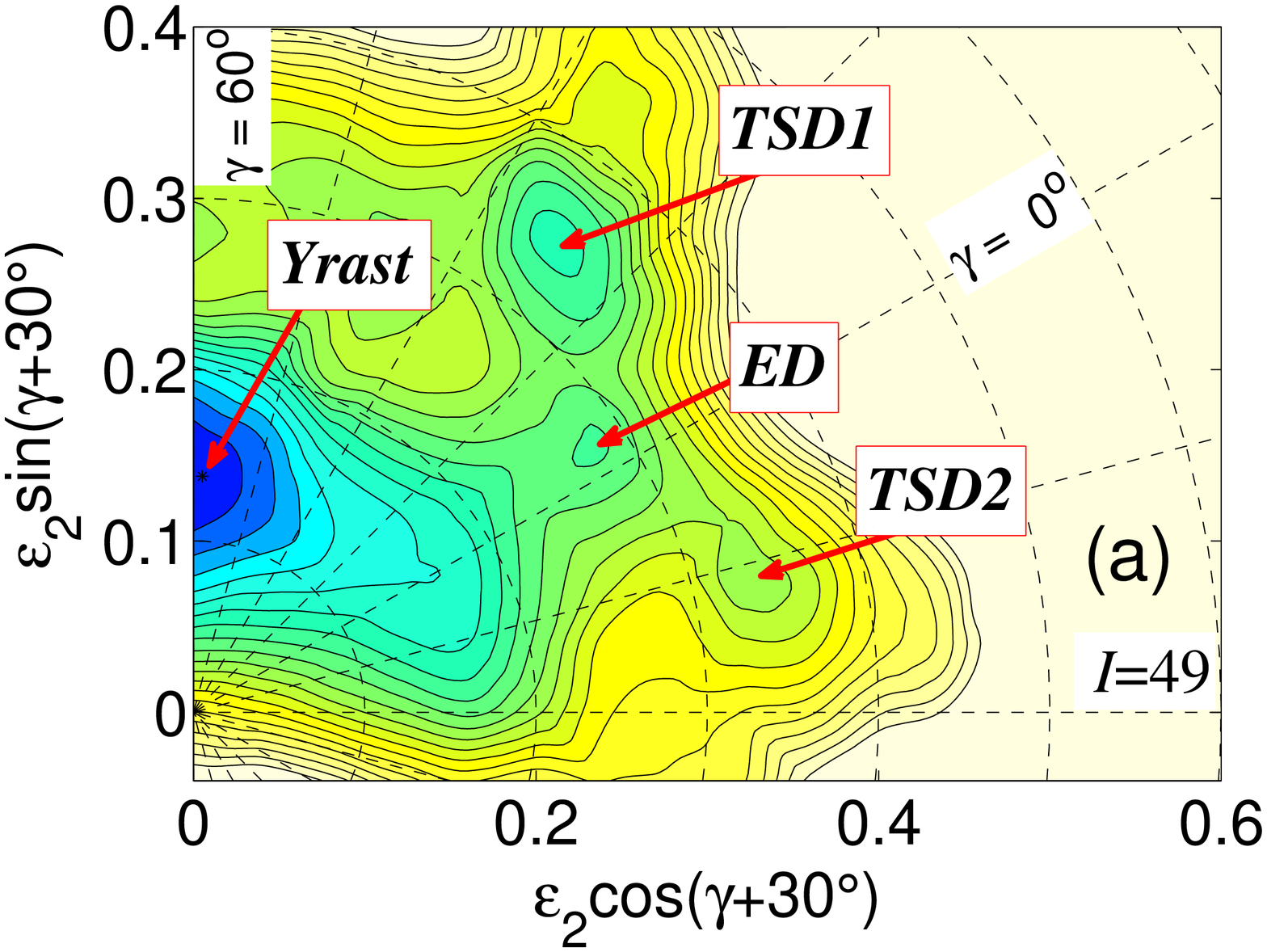}
\includegraphics[angle=0,width=0.86\columnwidth]{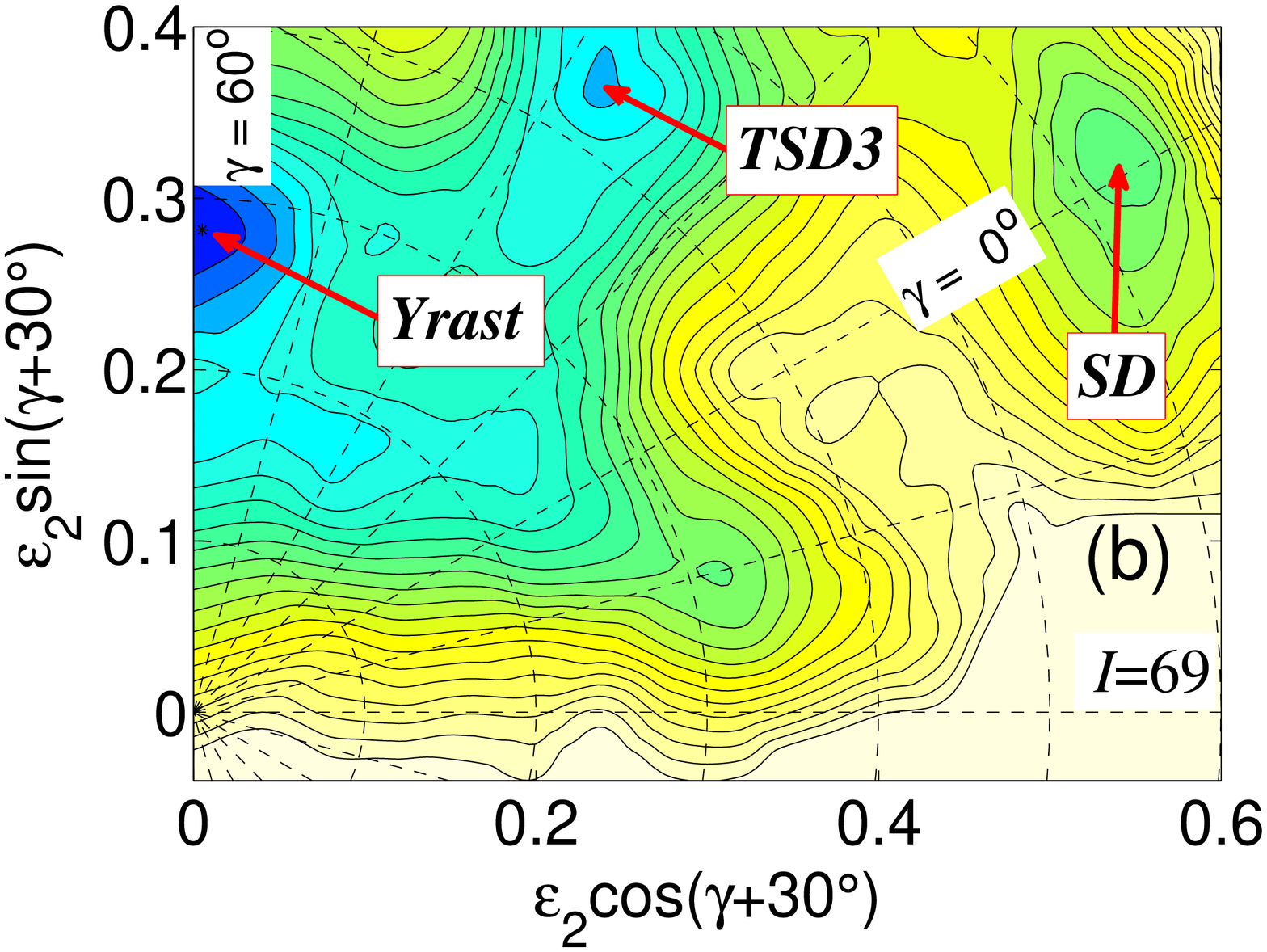}
\caption{(Color online) Potential energy surfaces in $^{158}$Er calculated for negative parity and signature $\alpha=1$ 
at spins $I=49$ (a) and 69 (b), respectively. The contour line separation is 0.25 MeV. The minima of 
interest are labelled (see text).\label{fig:158Er-PES}}
\end{center}
\end{figure}

\begin{figure}
\begin{center}
\includegraphics[angle=270,width=1.0\columnwidth]{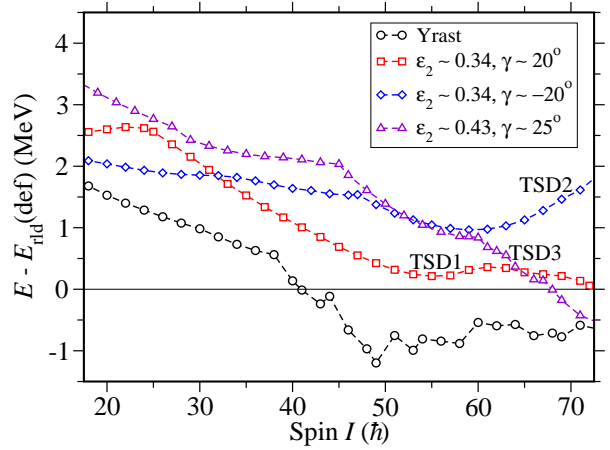}
\caption{(Color online) Calculated excitation energies of the lowest states associated with the three 
triaxial-shape minima and of the yrast states in $^{158}$Er plotted as a function of spin. 
The rotating liquid drop reference ($E_{\rm rld}$) was calculated using the 
approach of Ref.~\cite{Carlsson-PRC-74-R011302-06}.\label{fig:158Er-exer_ener_lowest}}
\end{center}
\end{figure}

In order to interpret these results, extensive CNS calculations without 
pairing~\cite{Carlsson-PRC-74-R011302-06} have been performed for $^{158}$Er. 
The potential energy surfaces (PES) for states with every combination of parity and signature exhibit similar 
energy minima as a function of spin. PES plots for negative parity and $\alpha=1$ signature (odd spin) 
at two representative spin values of 49 and $69{\hbar}$ are given in Fig.~\ref{fig:158Er-PES}(a) and 
\ref{fig:158Er-PES}(b), respectively. In Fig.~\ref{fig:158Er-PES}(a), there are four energy minima, 
one prolate, which is close to an axially symmetric shape with enhanced deformation at 
$\varepsilon_{2}\sim{0.28}$ (labelled as ED) when compared with the low-spin ($I{\le}30$) yrast structure 
with $\varepsilon_{2}\sim{0.2}$~\cite{Shepherd-PRC-65-034320-02}, a non-collective oblate minimum at $\varepsilon_{2}\sim{0.15}$ 
and $\gamma{=}60^{\circ}$ (labelled as Yrast), and two triaxial ones with $\varepsilon_{2}\sim{0.34}$, 
$\gamma{\sim}20^{\circ}$ (TSD1) and $\gamma{\sim}{-}20^{\circ}$ (TSD2). In Fig.~\ref{fig:158Er-PES}(b), 
the higher-spin PES plot, another more deformed triaxial minimum at $\varepsilon_{2}\sim{0.43}$ and 
$\gamma{\sim}25^{\circ}$ (TSD3) and a superdeformed one with $\varepsilon_{2}\sim{0.63}$ and 
$\gamma{\sim}0^{\circ}$ (SD) emerge together with a higher deformation oblate yrast structure. Of these minima, 
the non-collective oblate yrast minimum is associated with the terminating states~\cite{Tjom-PRL-55-2405-85,Simpson-PLB-327-187-94}, 
which are calculated lowest in energy in the spin range $40-70{\hbar}$ (see Fig.~\ref{fig:158Er-exer_ener_lowest}). 
The positive-$\gamma$ triaxial minimum TSD1 is the lowest energy strongly deformed minimum throughout the spin range 
$30-65{\hbar}$. Therefore, it was the favored candidate for interpreting the ultrahigh-spin 
structures observed in Ref.~\cite{Paul-PRL-98-012501-07}, prior to the present $Q_{\rm t}$ measurement. 

The general formula of $Q_{\rm t}$ for a triaxial nucleus~\cite{Juodagalvis-PLB-477-66-00} 
in terms of the density distribution along the three principal axes, 
$\hat{x}$ (rotation axis), $\hat{y}$, and $\hat{z}$ (symmetry axis for $\gamma{=}0^{\circ}$), is 
\[
Q_{\rm t}=\sqrt{\frac{8}{3}}|Q_{22}(\hat{x})|{\rm,~}
\]
where $Q_{22}(\hat{x})=\sqrt{\frac{3}{2}}{\langle}y^2-z^2{\rangle}$~\cite{Bohr-75-book}, 
which is equivalent to Eq. 15 in Ref.~\cite{Matev-PRC-76-034304-07}. 
The same radius parameter, $r_{0}=1.2~{\rm fm}$, has been employed 
as in previous studies of axial SD nuclei, see Ref.~\cite{Savajols-PRL-76-4480-96}, for example. 
In the $^{158}$Er case, the $Q_{\rm t}$ values calculated with this formula are $6.8-7.9~{\rm eb}$ for TSD1, 
$9.1-10.3~{\rm eb}$ for TSD2, and $8.6-9.9~{\rm eb}$ for TSD3. These energy minima migrate gradually in 
the ($\varepsilon_{2}$, $\gamma$) plane such that the associated $Q_{\rm t}$ value for each minimum decreases 
slightly with increasing spin and forms a range, illustrated by a shaded area in Fig.~\ref{fig:Qt_exp_cal_comp}. 
When calculating the $Q_{\rm t}$ value for a triaxial shape, it has often been 
assumed that $Q_{\rm t}$ depends on the $\gamma$ parameter as 
$Q_{\rm t}=|Q_{20}(\hat{z})|{\cos(\gamma+{30}^{\circ})}/{\cos(30^{\circ})}$~\cite{Ring-PLB-110-423-82}, 
where $Q_{20}(\hat{z})={\langle}2{z}^2-x^2-y^2{\rangle}$~\cite{Bohr-75-book} is the quadrupole moment of an axially deformed nucleus. 
The $Q_{\rm t}$ values for $^{158}$Er calculated with this simple, but approximate expression are smaller by ${\sim}8\%$, 
for the positive-$\gamma$ minima (TSD1 and TSD3), and larger by ${\sim}8\%$, for the negative-$\gamma$ minimum (TSD2), 
than those computed with the general formula above. However, these two formulae are equivalent for $\gamma{=}0^{\circ}$, 
and both give similar results for near-axially deformed minima, for example, $Q_{\rm t}{\sim}6.7-7.9~{\rm eb}$ 
for the ED case. 

The theoretical $Q_{\rm t}$ values associated with all collective minima, 
obtained with the general expression in the spin range $30-60{\hbar}$, 
are compared with the measured $Q_{\rm t}$ values of the $^{158}$Er bands 1 and 2 in Fig.~\ref{fig:Qt_exp_cal_comp}. 
It is clear that the experimental $Q_{\rm t}$ values, ${\sim}11~{\rm eb}$, are not consistent 
with either the calculated axially deformed minima, ED and SD, or the theoretically favored TSD1 minimum. 
The data are close to, but still larger, than the TSD2 and TSD3 triaxial minima. Without a clear preference, 
TSD2 still has calculated $Q_{\rm t}$ values slightly closer to the measured ones. However, 
TSD3 is predicted to be competitive with TSD2 in the $30-55{\hbar}$ spin range, and it is 
even lower than TSD2 for spins above $55{\hbar}$ (see Fig.~\ref{fig:158Er-exer_ener_lowest}). 
Therefore, when comparisons of $Q_{\rm t}$ are made with the CNS calculations, the ultrahigh-spin bands observed in 
$^{158}$Er appear most compatible with either the negative-$\gamma$ minimum TSD2 
($\varepsilon_{2}\sim{0.34}$) or the higher-deformed positive-$\gamma$ minimum TSD3 ($\varepsilon_{2}\sim{0.43}$). 

This is a surprising result, considering that the calculations consistently predict a 1 MeV gap between the TSD1 minimum 
and those associated with TSD2 and TSD3 in the $30-60{\hbar}$ spin range (Fig.~\ref{fig:158Er-exer_ener_lowest}). 
The observed bands appear to be the lowest in excitation energy since they are the only ones observed and 
the intensity profile indicates that they receive feeding from higher lying states over the entire $40-60{\hbar}$ spin range. 
The information on $Q_{\rm t}$ values, predicted minima, and feeding patterns for the two $^{157}$Er 
bands is very similar to that for $^{158}$Er. Therefore, the same general conclusion can be drawn in that there is an inconsistency between the 
observations and the previously favored theoretical interpretation~\cite{Paul-PRL-98-012501-07} for all the four bands observed in 
$^{157}$Er and $^{158}$Er. The lifetime data appear to rule out TSD1 as the relevant minimum and, while the calculated $Q_{\rm t}$ 
values for TSD2 and TSD3 are closer to the measured value for band 1, they are still outside the quoted errors. 

The present cranking calculations do not account for the $Q_{\rm t}$ data satisfactorily, although it is noted 
that these calculations also carry some uncertainties. 
It would be interesting to see what other theoretical approaches predict for the re-emergence of collectivity at these high 
spins in $^{157}$Er and $^{158}$Er. In the tilted-axis cranking (TAC) model~\cite{Frauendorf-NPA-677-115-00}, for example, 
it has been suggested that for a triaxial shape the rotational axis could tilt between the intermediate 
and short principal axes. Thus, a rotational band may be associated with a mixing of the positive- and negative-$\gamma$ 
minima~\cite{Pattabi-PLB-163Tm-07} and would have an average $Q_{\rm t}$ value between the positive- and negative-$\gamma$ 
limits. Perhaps, the present observations should be linked with the differences 
of $Q_{\rm t}$ values between theory and experiment for TSD bands and wobbling bands found in some Lu and Tm 
isotopes~\cite{Schonwa-EPJA-15-435-02,Wang-PRC-163Tm-07,Shimizu-PRC-77-024319-08}. While a satisfactory 
account can be given with CNS calculations for prolate superdeformed shapes~\cite{Savajols-PRL-76-4480-96}, 
the situation is different for bands which appear to be associated with triaxial strongly deformed shapes. On the other hand, the 
bands with large deformation in $^{168}$Hf have been described well~\cite{Kardan-to-be-published} in the CNS approach 
with a strongly deformed triaxial configuration. 
Obviously, a fully coherent understanding is yet to emerge. 

In summary, DSAM measurements demonstrate that the ultrahigh-spin sequences in $^{157,158}$Er are associated with 
strongly deformed shapes. The measured $Q_{\rm t}$ values appear to be more compatible, within the CNS theoretical framework, 
with a negative-$\gamma$ triaxial deformed minimum ($\varepsilon_{2}\sim{0.34}$) or a positive-$\gamma$ minimum with 
larger deformation ($\varepsilon_{2}{\sim}0.43$) rather than with the energetically favored positive-$\gamma$ triaxial shape 
($\varepsilon_{2}{\sim}0.34$) predicted by theory. Discrepancies, however, still remain between the experimental 
measurements and the theoretical predictions. This represents a challenge for the understanding of 
the triaxial degree of freedom in nuclei. 

The authors acknowledge Paul Morrall for preparing the targets, and the ATLAS operations staff for 
assistance. Discussions with S. Frauendorf and W. Nazarewicz are greatly appreciated. 
This work has been supported in part by the U.S. National Science Foundation under grants 
No. PHY-0756474 (FSU), PHY-0554762 (USNA), and PHY-0754674 (UND), the U.S. Department of Energy, Office of Nuclear Physics, 
under contracts No. DE-AC02-06CH11357 (ANL), DE-FG02-94ER40834 (UMD), DE-AC02-05CH11231 (LBL), and DE-FG02-96ER40983 (UTK), 
the United Kingdom Science and Technology Facilities Council, the Swedish Science Research Council, 
and by the State of Florida. 


\begin{thebibliography}{35}
\expandafter\ifx\csname natexlab\endcsname\relax\def\natexlab#1{#1}\fi
\providecommand{\bibinfo}[2]{#2}
\ifx\xfnm\relax \def\xfnm[#1]{\unskip,\space#1}\fi
\bibitem[{Bohr and Mottelson(1975)}]{Bohr-75-book}
\bibinfo{author}{A.~Bohr}, \bibinfo{author}{B.~R. Mottelson},
  \bibinfo{title}{Nuclear Structure}, volume~\bibinfo{volume}{II},
  \bibinfo{publisher}{Benjamin}, \bibinfo{address}{New York},
  \bibinfo{year}{1975}.
\bibitem[{{\O}deg{\aa}rd et~al.(2001){\O}deg{\aa}rd, Hagemann, Jensen,
  Bergstr{\"o}m, Herskind, Sletten, T{\"o}rm{\"a}nen, Wilson, Tj{\o}m,
  Hamamoto, Spohr, H{\"u}bel, G{\"o}rgen, Sch{\"o}nwasser, Bracco, Leoni, Maj,
  Petrache, Bednarczyk, and Curien}]{Odegard-PRL-86-5866-01}
\bibinfo{author}{S.~W. {\O}deg{\aa}rd $\it{et~al.}$},
  \bibinfo{journal}{Phys. Rev. Lett.} \bibinfo{volume}{86}
  (\bibinfo{year}{2001}) \bibinfo{pages}{5866}.
\bibitem[{Andersson et~al.(1976)Andersson, Larsson, Leander, M{\"o}ller,
  Nilsson, Ragnarsson, {\AA}berg, Bengtsson, Dudek, Nerlo-Pomorska, Pomorski,
  and Szymanski}]{Andersson-NPA-268-205-76}
\bibinfo{author}{G.~Andersson $\it{et~al.}$},
  \bibinfo{journal}{Nucl. Phys. A} \bibinfo{volume}{268} (\bibinfo{year}{1976})
  \bibinfo{pages}{205}.
\bibitem[{Szyma\'{n}ski(1983)}]{Szymanski-83-book}
\bibinfo{author}{Z.~Szyma\'{n}ski}, \bibinfo{title}{Fast Nuclear Rotation},
  \bibinfo{publisher}{Clarendon Press}, \bibinfo{address}{Oxford, England},
  \bibinfo{year}{1983}.
\bibitem[{Carlsson et~al.(2008)Carlsson, Ragnarsson, Bengtsson, Lieder, Lieder,
  and Pasternak}]{Carlsson-PRC-78-034316-08}
\bibinfo{author}{B.~G. Carlsson $\it{et~al.}$},
  \bibinfo{journal}{Phys. Rev. C} \bibinfo{volume}{78} (\bibinfo{year}{2008})
  \bibinfo{pages}{034316}.
\bibitem[{Paul et~al.(2007)Paul, Twin, Evans, Pipidis, Riley, Simpson, Appelbe,
  Campbell, Choy, Clark, Cromaz, Fallon, G{\"o}rgen, Joss, Lee, Macchiavelli,
  Nolan, Ward, and Ragnarsson}]{Paul-PRL-98-012501-07}
\bibinfo{author}{E.~S. Paul $\it{et~al.}$},
  \bibinfo{journal}{Phys. Rev. Lett.} \bibinfo{volume}{98} (\bibinfo{year}{2007})
  \bibinfo{pages}{012501}.
\bibitem[{Tj\o{}m et~al.(1985)Tj\o{}m, Diamond, Bacelar, Beck, Deleplanque,
  Draper, and Stephens}]{Tjom-PRL-55-2405-85}
\bibinfo{author}{P.~O. Tj\o{}m $\it{et~al.}$},
  \bibinfo{journal}{Phys. Rev. Lett.} \bibinfo{volume}{55} (\bibinfo{year}{1985})
  \bibinfo{pages}{2405}.
\bibitem[{Simpson et~al.(1994)Simpson, Riley, Gale, Sharpey-Schafer, Bentley,
  Bruce, Chapman, Clark, Clarke, Copnell, Cullen, Fallon, Fitzpatrick, Forsyth,
  Freeman, Jones, Joyce, Liden, Lisle, Macchiavelli, Smith, Smith, Sweeney,
  Thompson, Warburton, Wilson, Bengtsson, and
  Ragnarsson}]{Simpson-PLB-327-187-94}
\bibinfo{author}{J.~Simpson $\it{et~al.}$},
  \bibinfo{journal}{Phys. Lett. B} \bibinfo{volume}{327} (\bibinfo{year}{1994})
  \bibinfo{pages}{187}.
\bibitem[{Bengtsson and Ragnarsson(1983)}]{Bengtsson-PSC-T5-165-83}
\bibinfo{author}{T.~Bengtsson}, \bibinfo{author}{I.~Ragnarsson},
  \bibinfo{journal}{Phys. Scr.} \bibinfo{volume}{T5} (\bibinfo{year}{1983})
  \bibinfo{pages}{165}.
\bibitem[{Dudek and Nazarewicz(1985)}]{Dudek-PRC-31-298-85}
\bibinfo{author}{J.~Dudek}, \bibinfo{author}{W.~Nazarewicz},
  \bibinfo{journal}{Phys. Rev. C} \bibinfo{volume}{31} (\bibinfo{year}{1985})
  \bibinfo{pages}{298}.
\bibitem[{Carlsson and Ragnarsson(2006)}]{Carlsson-PRC-74-R011302-06}
\bibinfo{author}{B.~G. Carlsson}, \bibinfo{author}{I.~Ragnarsson},
  \bibinfo{journal}{Phys. Rev. C} \bibinfo{volume}{74} (\bibinfo{year}{2006})
  \bibinfo{pages}{011302(R)}.
\bibitem[{Devons et~al.(1955)Devons, Manning, and
  Bunbury}]{Devons-PPSA-68-18-55}
\bibinfo{author}{S.~Devons}, \bibinfo{author}{G.~Manning},
  \bibinfo{author}{D.~S.~P. Bunbury}, \bibinfo{journal}{Proc. Phys. Soc. A}
  \bibinfo{volume}{68} (\bibinfo{year}{1955}) \bibinfo{pages}{18}.
\bibitem[{Janssens and Stephens(1996)}]{Janssens-NPN-6-9-96}
\bibinfo{author}{R.~V.~F. Janssens}, \bibinfo{author}{F.~S. Stephens},
  \bibinfo{journal}{Nucl. Phys. News} \bibinfo{volume}{6}
  (\bibinfo{year}{1996}) \bibinfo{pages}{9}.
\bibitem[{Fischer et~al.(1996)Fischer, Janssens, Riley, Chasman, Ahmad,
  Blumenthal, Brown, Carpenter, Hackman, Hartley, Khoo, Lauritsen, Ma, Nisius,
  Simpson, and Varmette}]{Fischer-PRC-54-R2806-96}
\bibinfo{author}{S.~M. Fischer $\it{et~al.}$},
  \bibinfo{journal}{Phys. Rev. C} \bibinfo{volume}{54} (\bibinfo{year}{1996})
  \bibinfo{pages}{2806(R)}.
\bibitem[{Wang et~al.(2007)Wang, Janssens, Moore, Garg, Gu, Frauendorf,
  Carpenter, Ghugre, Hammond, Lauritsen, Li, Mukherjee, Pattabiraman,
  Seweryniak, and Zhu}]{Wang-PRC-163Tm-07}
\bibinfo{author}{X.~Wang $\it{et~al.}$},
  \bibinfo{journal}{Phys. Rev. C} \bibinfo{volume}{75} (\bibinfo{year}{2007})
  \bibinfo{pages}{064315}.
\bibitem[{Radford(1995)}]{Radford-NIMA-361-306-95}
\bibinfo{author}{D.~C. Radford}, \bibinfo{journal}{Nucl. Instr. And Meth. A}
  \bibinfo{volume}{361} (\bibinfo{year}{1995}) \bibinfo{pages}{306}.
\bibitem[{Cromaz et~al.(2001)Cromaz, Symons, Lane, Lee, and
  MacLeod}]{Cromaz-NIMA-462-519-01}
\bibinfo{author}{M.~Cromaz $\it{et~al.}$}, 
  \bibinfo{journal}{Nucl. Instr. And Meth. A} \bibinfo{volume}{462} (\bibinfo{year}{2001})
  \bibinfo{pages}{519}.
\bibitem[{Starosta et~al.(2003)Starosta, Fossan, Koike, Vaman, Radford, and
  Chiara}]{Starosta-NIMA-515-771-03}
\bibinfo{author}{K.~Starosta $\it{et~al.}$},
  \bibinfo{journal}{Nucl. Instr. And Meth. A} \bibinfo{volume}{515}
  (\bibinfo{year}{2003}) \bibinfo{pages}{771}.
\bibitem[{Moore(????)}]{Moore-unpublsh}
\bibinfo{author}{E.~F. Moore}, \bibinfo{note}{private communication}.
\bibitem[{Ziegler et~al.(1985)Ziegler, Biersack, and
  Littmark}]{Ziegler-85-book}
\bibinfo{author}{J.~F. Ziegler}, \bibinfo{author}{J.~P. Biersack},
  \bibinfo{author}{U.~Littmark}, \bibinfo{title}{The stopping and Range of Ions
  in Solids}, \bibinfo{publisher}{Pergamon}, \bibinfo{address}{New York},
  \bibinfo{year}{1985}.
\bibitem[{Nisius et~al.(1997)Nisius, Janssens, Moore, Fallon, Crowell,
  Lauritsen, Hackman, Ahmad, Amro, Asztalos, Carpenter, Chowdhury, Clark, Daly,
  Deleplanque, Diamond, Fischer, Grabowski, Khoo, Lee, Macchiavelli, Mayer,
  Stephens, Afanasjev, and Ragnarsson}]{Nisius-PLB-392-18-97}
\bibinfo{author}{D.~Nisius $\it{et~al.}$},
  \bibinfo{journal}{Phys. Lett. B} \bibinfo{volume}{392} (\bibinfo{year}{1997})
  \bibinfo{pages}{18}.
\bibitem[{Amro et~al.(2001)Amro, Varmette, Ma, Herskind, Hagemann, Sletten,
  Janssens, Bergstr{\"o}m, Bracco, Carpenter, Domscheit, Frattini, Hartley,
  H{\"u}bel, Khoo, Kondev, Lauritsen, Lister, Million, {\O}deg{\aa}rd, Piercey,
  Riedinger, Schmidt, Siem, Wiedenh{\"o}ver, Wilson, and
  Winger}]{Amro-PLB-506-39-01}
\bibinfo{author}{H.~Amro $\it{et~al.}$},
  \bibinfo{journal}{Phys. Lett. B} \bibinfo{volume}{506} (\bibinfo{year}{2001})
  \bibinfo{pages}{39}.
\bibitem[{Wang(2007)}]{Wang-07-thesis}
\bibinfo{author}{X.~Wang}, Ph.D. thesis, University of Notre Dame,
  \bibinfo{year}{2007}. \bibinfo{note}{ArXiv:0810.5021}.
\bibitem[{Shepherd et~al.(2002)Shepherd, Simpson, Dewald, Petkov, Nolan, Riley,
  Boston, Brown, Clark, Fallon, Hartley, Kasemann, Kr\"ucken, von Brentano,
  Laird, Paul, and Peusquens}]{Shepherd-PRC-65-034320-02}
\bibinfo{author}{S.~L. Shepherd $\it{et~al.}$},
  \bibinfo{journal}{Phys. Rev. C} \bibinfo{volume}{65} (\bibinfo{year}{2002})
  \bibinfo{pages}{034320}.
\bibitem[{Broude(1973)}]{Broude-73-book}
\bibinfo{author}{C.~Broude}, \bibinfo{title}{Lecture Notes in Physics: Stopping
  power effects in nuclear lifetime measurements}, volume~\bibinfo{volume}{23},
  \bibinfo{publisher}{Springer Berlin / Heidelberg}, \bibinfo{year}{1973}.
\bibitem[{Revill(????)}]{Revill-to-be-published}
\bibinfo{author}{J.~P. Revill $\it{et~al.}$}, \bibinfo{note}{to be published}.
\bibitem[{Juodagalvis et~al.(2000)Juodagalvis, Ragnarsson, and
  {\AA}berg}]{Juodagalvis-PLB-477-66-00}
\bibinfo{author}{A.~Juodagalvis}, \bibinfo{author}{I.~Ragnarsson},
  \bibinfo{author}{S.~{\AA}berg}, \bibinfo{journal}{Phys. Lett. B}
  \bibinfo{volume}{477} (\bibinfo{year}{2000}) \bibinfo{pages}{66}.
\bibitem[{Matev et~al.(2007)Matev, Afanasjev, Dobaczewski, Lalazissis, and
  Nazarewicz}]{Matev-PRC-76-034304-07}
\bibinfo{author}{M.~Matev $\it{et~al.}$},
  \bibinfo{journal}{Phys. Rev. C} \bibinfo{volume}{76} (\bibinfo{year}{2007})
  \bibinfo{pages}{034304}.
\bibitem[{Savajols et~al.(1996)Savajols, Korichi, Ward, Appelbe, Ball,
  Beausang, Beck, Byrski, Curien, Dagnall, de~France, Disdier, Duch\^ene,
  Erturk, Finck, Flibotte, Gall, Galindo-Uribarri, Haas, Hackman, Janzen,
  Kharraja, Lisle, Merdinger, Mullins, Pilotte, Pr\'evost, Radford, Rauch,
  Rigollet, Smalley, Smith, Stezowski, Styczen, Theisen, Twin, Vivien,
  Waddington, Zuber, and Ragnarsson}]{Savajols-PRL-76-4480-96}
\bibinfo{author}{H.~Savajols $\it{et~al.}$},
  \bibinfo{journal}{Phys. Rev. Lett.} \bibinfo{volume}{76}
  (\bibinfo{year}{1996}) \bibinfo{pages}{4480}.
\bibitem[{Ring et~al.(1982)Ring, Hayashi, Hara, Emling, and
  Grosse}]{Ring-PLB-110-423-82}
\bibinfo{author}{P.~Ring $\it{et~al.}$}, 
  \bibinfo{journal}{Phys. Lett. B} \bibinfo{volume}{110} (\bibinfo{year}{1982})
  \bibinfo{pages}{423}.
\bibitem[{Frauendorf(2000)}]{Frauendorf-NPA-677-115-00}
\bibinfo{author}{S.~Frauendorf}, \bibinfo{journal}{Nucl. Phys. A}
  \bibinfo{volume}{677} (\bibinfo{year}{2000}) \bibinfo{pages}{115}.
\bibitem[{Pattabiraman et~al.(2007)Pattabiraman, Gu, Frauendorf, Garg, Li,
  Nayak, Wang, Zhu, Ghugre, Janssens, Chakrawarthy, Whitehead, and
  Macchiavelli}]{Pattabi-PLB-163Tm-07}
\bibinfo{author}{N.~S. Pattabiraman $\it{et~al.}$}, 
  \bibinfo{journal}{Phys. Lett. B} \bibinfo{volume}{647}
  (\bibinfo{year}{2007}) \bibinfo{pages}{243}.
\bibitem[{Sch{\"{o}}nwa{\ss}er et~al.(2002)Sch{\"{o}}nwa{\ss}er, H{\"u}bel,
  Hagemann, Amro, Clark, Cromaz, Diamond, Fallon, Herskind, Lane, Ma,
  Macchiavelli, {\O}deg{\aa}rd, Sletten, Ward, and
  Wilson}]{Schonwa-EPJA-15-435-02}
\bibinfo{author}{G.~Sch{\"{o}}nwa{\ss}er $\it{et~al.}$},
  \bibinfo{journal}{Eur. Phys. J. A} \bibinfo{volume}{15}
  (\bibinfo{year}{2002}) \bibinfo{pages}{435}.
\bibitem[{Shimizu et~al.(2008)Shimizu, Shoji, and
  Matsuzaki}]{Shimizu-PRC-77-024319-08}
\bibinfo{author}{Y.~R. Shimizu}, \bibinfo{author}{T.~Shoji},
  \bibinfo{author}{M.~Matsuzaki}, \bibinfo{journal}{Phys. Rev. C}
  \bibinfo{volume}{77} (\bibinfo{year}{2008}) \bibinfo{pages}{024319}.
\bibitem[{Kardan and Ragnarsson(????)}]{Kardan-to-be-published}
\bibinfo{author}{A.~Kardan}, \bibinfo{author}{I.~Ragnarsson}, \bibinfo{note}{to be published}.

\end{thebibliography}

\end{document}